\documentstyle[aps,prl,twocolumn]{revtex}

\newcommand{\Tr}{{\,\rm Tr\,}}

\draft
\input psfig.sty
\begin {document}

\title {Dual description of nonextensive ensembles}
\author {Jan Naudts}
\address {Departement Natuurkunde, Universiteit Antwerpen,
          UIA, 2610 Antwerpen, Belgium}
%\date {}
\maketitle

\begin {abstract}

\noindent
Consider a quantum mechanical system with an unbounded hamiltonian. For
such a system the entropic index $q$ of non-extensive thermodynamics has
an upperbound $q_c\ge 1$ beyond which the formalism becomes meaningless.
The expression $1/(q_c-1)$ is the dimension of the state space (i.e.~the
manifold of density matrices) in the context of non-commutative
geometry. For $q=q_c$ an ultraviolet cutoff $E\le E_{\rm ub}$ is needed
to guarantee the existence of the equilibrium density matrix. Duality
between $q>1$ and $q<1$-statistics via a $q\rightarrow 1/q$
transformation is established. It leads to an overall picture in which
the meaning of both $q>1$-statistics and $q<1$-statistics is clarified.
The hydrogen atom is considered as an example.


\end {abstract}


\bigskip
The aim of this paper is to contribute to the physical understanding of
nonextensive thermodynamics. It is obvious from the many applications
(see e.g.~\cite {TC99}) that the formalism is physically relevant.
Nevertheless the overall picture is still unclear. The open question
treated below is wether physical systems should be described by an
entropic index $q$ which is less than 1 or larger than 1. The
$q>1$-situation is easier to handle. On the other hand there are
arguments in favor of $q<1$. Both views are reconciled by the
$q\rightarrow 1/q$-duality which was proposed in \cite {TMP98}. By
working out this duality it becomes feasible to attach a physical
meaning to both $q<1$ and $q>1$ formalisms.

%intrinsic q-value
In \cite {TMP98} the opinion is expressed that each hamiltonian has an
intrinsic $q$-value, and that systems with short range interactions all
belong to the universality class of $q=1$-systems. Such an intrinsic
value, denoted $q_c$, is proposed here. At first sight it does not seem
to agree with the statement about the universality class of systems with
short range interactions. However, at the end of the paper we will see
that a hamiltonian with long range potential, like that of the hydrogen
atom, does indeed determine a $q_c$-value different from 1.

% intro linking two fields
A side effect of the present work is a link with noncommutative
geometry as formulated in \cite {CA94}.
This mathematical theory generalizes concepts
of differential geometry to quantum mechanics and could be essential in
reconciling quantum mechanics with theories of gravity. An introductory
text is \cite {LG97}. An essential result of
this theory is that in a non-commutative context the geometry of the
manifold of states is fully determined by an operator which in
in the case of relativistic quantum mechanics is the Dirac operator.
This correspondence is used
here to interprete the expression $1/(q_c-1)$ as a geometric dimension
At this moment it is not clear wether this correspondence will lead to
new insights. It is however not excluded that non-extensivity plays an
important role in the quantum mechanics of the microcosmos, as is also
suggested by previous links between non-extensive thermodynamics and
quantum groups \cite {TC94},\cite {AKY97},
resp.~q-deformations {AS97}.

% entropic parameter $q$, Tsallis thermodyn
The Tsallis entropy is given by \cite {TC88}
\begin {equation}
{\cal S}_q(\rho)=k_B{1-\Tr\rho^q\over q-1}
\end {equation}
It depends on the entropic parameter $q\not=1$
(we assume throughout the paper that $q>0$)
and on the density matrix $\rho$
($\rho\ge 0$ and $\Tr\rho=1$).
In the limit $q\rightarrow 1$ the usual
Shannon entropy is recovered. Minimization
of the free energy
\begin {equation}
{\cal F}_q={\cal U}^H_q-T{\cal S}_q
\end {equation}
determines the equilibrium density matrix at
temperature $T$. For the energy ${\cal U}^H_q$
several proposals have been made, the latest
of which is \cite {TMP98}
\begin {equation}
{\cal U}^H_q(\rho)={\Tr\rho^q H\over \Tr\rho^q}
\label {energ}
\end {equation}
The hamiltonian $H$ has a discrete spectrum bounded from below. The
inverse temperature will be denoted $\beta=1/k_BT$ as usual.

% q>1 story first
Different situations occur wether $q>1$ or $q<1$. Let us first assume
that $q>1$. Because of the duality, which will be discussed later on, the
equilibrium density matrix in the $q>1$-case is denoted $\sigma$ instead
of $\rho$. For the same reason the hamiltonian is denoted $K$ instead of
$H$. The equilibrium density matrix is of the form
\begin {equation}
\sigma={1\over\zeta} \left[{\bf 1}
-\beta (1-q){K-{\cal U}^K_q(\sigma){\bf 1}
\over \Tr\sigma^q}\right]^{1/(1-q)}
\label {qmo}
\end {equation}
with
\begin {equation}
\zeta=\Tr \left[{\bf 1}
-\beta (1-q){K-{\cal U}^K_q(\sigma){\bf 1}
\over \Tr\sigma^q}\right]^{1/(1-q)}
\label {zetaeq}
\end {equation}
Note that Eq.~(\ref {qmo}) is of the implicit type.
One can prove that it has a
unique solution, provided that the operator
$K^{1/(1-q)}$ is trace class (see \cite {NC99}).
Let us consider the latter condition in more detail.
Let $(\psi_n)_n$ be a basis in which $K$ is diagonal,
with eigenvalues $\epsilon_n$: $K\psi_n=\epsilon_n\psi_n$.
Assume that the eigenvalues are isolated, with finite multiplicity,
and ordered increasingly.
Let $q_c$ be the lower bound of $q\ge 1$
for which the sequence $n^{1-q}\epsilon_n$ is bounded.
Then obviously the eigenvalues of $K$ diverge as
\begin {equation}
\epsilon_n\sim n^{q_c-1}
\end {equation}
Note that the dimension of the non-commutative state space equals
$1/(q_c-1)$. See e.g.~\cite {LG97}, section 5.6.
Let us consider some examples.
If $K$ were bounded (which is assumed not to be the case)
then $q_c$ would be 1. For the 
harmonic oscillator $q_c$ equals 2, which corresponds
with a geometric dimension equal to 1. For a quantum particle in a
d-dimensional box with reflecting walls $q_c=1+2/d$.
Hence the dimension of the manifold is $d/2$, which is
only half of what is expected. The explanation is that the
laplacean is a second order operator. The correct
dimension is obtained when the Dirac operator or the
square root of minus the laplacean is taken as the operator
determining the geometry of the state space.

The critical entropic index $q_c$ is used as follows.
If $q<q_c$ then $K^{1/(1-q)}$ is trace class
while if $q>q_c$ it is certainly not. Hence $q_c$
is the upperbound of the entropic index
for which the nonextensive thermodynamics can still be
formulated (assuming energy has the form (\ref {energ})).

If $q=q_c$ then the trace in (\ref {zetaeq}) diverges generically.
Finite values can be obtained by introducing
a cutoff in energy, i.e.~by fixing a large number $N$,
and restricting the system to the finite Hilbert space 
spanned by the eigenvectors $\psi_n,n=0, 1,\cdots,N-1$.
Note that the cutoff in energy $E\le E_{\rm ub}$
is related to $N$ by
\begin {equation}
E_{\rm ub}\sim N^{q_c-1}
\end {equation}
In the limit of large $N$ the trace of (\ref {zetaeq})
can be estimated in terms of the Dixmier trace $\Tr_\omega$,
which is defined by
\begin {equation}
\Tr_\omega A=\lim_{N\rightarrow\infty}{1\over\ln N}\sum_{n=0}^{N-1}
\langle\psi_n|A|\psi_n\rangle
\qquad\hbox{ for all }A
\end {equation}
The expectation value of any
observable $A$ is then given by
\begin {equation}
\langle A\rangle
=\lim_{N\rightarrow\infty}\Tr_N\sigma_N A
={
\Tr_\omega {a{\bf 1}+K}^{1/(1-q_c)}A
\over
\Tr_\omega {a{\bf 1}+K}^{1/(1-q_c)}
}
\label {integral}
\end  {equation}
with $a$ an arbitrary constant such that
$a{\bf 1}+K>0$.
Remarkable is that $\langle A\rangle$ does not depend on $\beta$.
Its value is determined by the high energy tail of
the spectrum of $K$.
In the context of noncommutative geometry
(\ref {integral}) is called the integral of $A$.

The $q$-averages are given by
\begin {equation}
\langle A\rangle_q=\Tr\rho_N A
={\Tr\sigma_N^{q_c}A\over \Tr \sigma_N^{q_c}}
\end  {equation}
These are the physical expectation values.
They depend on temperature and cutoff via
a rescaled inverse temperature
\begin {equation}
\beta^*=\beta(\ln N)^{q_c+1}
\end  {equation}
A straightforward calculation shows that
\begin {equation}
\langle A\rangle_q
\simeq
{\Tr(f(\beta^*)+K_N)^{q_c/(1-q_c)}A
\over \Tr (f(\beta^*)+K)^{q_c/(1-q_c)}}
\end  {equation}
with $f$ the inverse of the function $g$ given by
\begin {equation}
g(a)={\left( \Tr(a+K)^{q_c/(1-q_c)}\right)^2
\over
\left(\Tr_\omega (a+K)^{1/(1-q_c)}\right)^{q+1}
}
\label {gfun}
\end  {equation}

%0<q<1
Let us now consider the case $0<q<1$.
The equilibrium density matrix is the
solution of the implicit equation
\begin {equation}
\rho={1\over\zeta} \left[{\bf 1}
-\beta (1-q){H-{\cal U}^H_q(\rho){\bf 1}\over \Tr\rho^q}\right]_+^{1/(1-q)}
\label {qlo}
\end {equation}
with $\displaystyle\left[A\right]_+=A$
on the subspace on which $A$ is positive, and $=0$
on the subspace on which $A$ is negative,
and with
\begin {equation}
\zeta=\Tr \left[{\bf 1}
-\beta (1-q){H-{\cal U}^H_q(\rho){\bf 1}\over \Tr\rho^q}\right]_+^{1/(1-q)}
\end {equation}
The proof that under certain conditions these equations determine
a unique equilibrium density matrix is complicated by the presence
of the high energy cutoff --- see {\cite {NC99}} for a full discussion;
part of the problem is discussed below.

% finite energy - plastino heath bath
In \cite {PP94} it was argued that a system in thermal contact with
a finite heath bath is described by Tsallis thermodynamics with
entropic index $0<q<1$. In this context the appearance of an energy
cutoff is very natural since the total energy of system plus heath
bath is finite. This bound on the total energy belongs probably to the
essence of nonextensive systems.

%duality
A $q\sim 1/q$-duality  has been studied in \cite {TMP98}.
It is based on the observation that the relation
$\rho=\sigma^q/\Tr\sigma^q$
can be inverted to
$\sigma=\rho^{q'}/\Tr\rho^{q'}$
with $q'=1/q$.
This duality is worked out below by introduction of
the effective Hamiltonian $H_{\rm eff}$ of \cite {CN99}.
Write the
energy functional (\ref {energ}) in a 1-homogeneous form
\begin {equation}
{\cal U}^H_q(\rho)=(\Tr\rho){\Tr\rho^q H\over \Tr\rho^q}
\label {energ2}
\end {equation}
Then the effective Hamiltonian $H_{\rm eff}$ is defined
by variation of the energy functional
\begin {equation}
H_{\rm eff}(\rho)={\delta\ \over\delta\rho}{\cal U}^H_q(\rho)
\end {equation}
(the variation is restricted to the subspace on which
$\rho$ is strictly positive).
It is the generator of the time evolution (see \cite {CN99}).
For an equilibrium state $[\rho,H]=0$ holds, in which
case it is easy to obtain (putting $\Tr\rho=1$)
\begin {equation}
H_{\rm eff}(\rho)={\cal U}^H_q(\rho){\bf 1}
+{q\over\Tr\rho^q}\rho^{q-1}
\left(H-{\cal U}^H_q(\rho){\bf 1}\right)
\end {equation}
Note also that
\begin {equation}
\Tr\sigma H=\Tr\rho H_{\rm eff}(\rho)={\cal U}^H_q(\rho)
\end {equation}
because of the 1-homogeneity of (\ref {energ2}).
A short calculation now gives
\begin {eqnarray}
& &{\bf 1}-\beta(1-q){H-{\cal U}^H_q(\rho){\bf 1}
\over \Tr\rho^q}\cr
& &\quad=\left[
{\bf 1}-\beta(1-{1\over q})\lambda
{H_{\rm eff}(\rho)-{\cal U}^H_q(\rho){\bf 1}
\over \Tr\sigma^{1/q}}
\right]^{-1}
\label {dualrel}
\end {eqnarray}
with
$\lambda={\Tr\sigma^{1/q}/\Tr\rho^q}$.
The lhs of (\ref {dualrel}) is proportional to
$\rho^{1-q}$ on the subspace on which it is strictly
positive. Hence from $\sigma\sim \rho^q$ follows
\begin {equation}
\sigma={1\over\xi}
\left[
{\bf 1}-\beta(1-{1\over q})\lambda{
H_{\rm eff}(\rho)-{\cal U}^H_q(\rho){\bf 1}
\over \Tr\sigma^{1/q}}
\right]^{1/(1-1/q)}
\label {sigsol}
\end {equation}
with $\xi$ the normalization factor. It turns out that
\begin {equation}
\xi=\zeta^q\Tr\rho^q
\end {equation}
There is no cutoff in (\ref {sigsol}) because
$H_{\rm eff}(\rho)$ is zero on the subspace where
the cutoff is active.
(\ref {sigsol}) shows that $\sigma$
is the equilibrium density matrix corresponding
to the entropic index $1/q$ for a system with
hamiltonian $K=\lambda H_{\rm eff}(\rho)$.
One concludes that a nice duality between $q<1$
and $q>1$-statistics is obtained
if not only $q$ is mapped onto $1/q$ and $\rho$ is mapped onto
$\sigma$, but also $H$ is mapped onto $K=\lambda H_{\rm eff}(\rho)$.

%consequences of duality
From this duality the following statements can be deduced.
\begin {itemize}

\item {} $q>1$-statistics is another way of looking at
$q<1$-statistics. The population of energy levels as described by
the $q<1$-density matrix $\rho$ drops faster than exponentially
with increasing energy, as it is expected in a non-extensive
system in thermal contact with a finite heath bath. The density
matrix $\sigma$ of $q>1$-statistics is a renormalized density
matrix in which the population of high energy levels has
been increased in an artificial way.
\item {}
Not every system described with $q>1$-statistics can be obtained
in this way.
Indeed, due to the cutoff in (\ref {qlo}) the effective hamiltonian
$H_{\rm eff}$ is in most cases defined on a finite dimensional Hilbert space.
Hence a cutoff should be applied to the $q>1$-hamiltonian $K$ before
considering it as an effective hamiltonian. Further complications
are that the effective hamiltonian depends on temperature, and
that it is {\sl not} the result of a simple ultraviolet cutoff
applied to a fixed hamiltonian.
\item {} Two types of hamiltonians are involved. The hamiltonian
$H$ in the $q<1$-description is the operator which determines
the energy of the system. The operator $K$ of the
$q>1$-description corresponds with the effective hamiltonian
$H_{\rm eff}$ of
the $q<1$-case and hence is also the generator of the dynamics.
This observation is in agreement with the use of $K$ in the
context of noncommutative geometry to formulate
Tomita-Takesaki-theory (see \cite {LG97}, section 5.7).
The connection with linear respons theory and the KMS condition
will be discussed elsewhere \cite {NJ99}.
\end {itemize}

%inverse duality transformation
Let us now consider which kind of system at $q<1$
corresponds by duality with a system with entropic
index $q$ satisfying $1<q<q_c$.
For most hamiltonians the equilibrium density
matrix of the $0<q<1$-case will contain only a finite number
of eigenvalues different from zero. This is a consequence of the
high-energy cutoff in (\ref {qlo}). However, there is one exception,
when the hamiltonian $H$ is bounded and
has an accumulation of eigenvalues at
the upper limit $\epsilon_\infty$ of its spectrum. This is
e.g.~the case with the discrete part of the spectrum of the hydrogen
atom. This is precisely the kind of hamiltonian for which
$q<1$-statistics is physically meaningful because the usual Gibbs
statistics cannot be applied - see \cite {LST95}.
Let $\beta_-$ be the solution of the equation
\begin {equation}
\Tr\rho^q=\beta(1-q)(\epsilon_\infty-{\cal U}_q^H(\rho))
\end {equation}
If $\beta>\beta_-$ then the energy cutoff is active and only a finite
number of eigenvalues of the equilibrium density matrix differ from
zero. At high temperatures, i.e.~$\beta<\beta_-$, the operator in (\ref
{qlo}) has a diverging trace and $q$-statistics is meaningless, as it is
the case for $q=1$. The inverse temperature $\beta_-$ depends on $q$. It
tends to $\infty$ when $q\rightarrow 1$ and when $q\rightarrow 1/q_c$,
with the critical entropic index $q_c$ equal to the upperbound of $q$'s
for which $n^{q-1}(\epsilon_\infty-\epsilon_n)$ is bounded,
i.e.
\begin {equation}
\epsilon_\infty-\epsilon_n\sim {1\over n^{q_c-1}}
\end {equation}
One concludes that the thermodynamic equilibrium of $q<1$-statistics
exists for $q$ in the interval $(1/q_c,1)$ and for low enough
temperatures so that $\beta >\beta_-$. For completeness, note that
for a bounded hamiltonian in an infinite dimensional Hilbert space
the free energy has no minimum. Indeed, the energy ${\cal U}_q(\rho)$
is bounded above, while the entropy ${\cal S}_q(\rho)$ can be made
arbitrary large. Hence the free energy ${\cal F}_q(\rho)$ can be made
arbitrary small. The density matrix given by (\ref {qlo}) is only
a relative minimum of the free energy. However it is the unique
maximum of the entropy ${\cal S}_q(\rho)$ for a given value of
the energy ${\cal U}_q(\rho)$ --- see \cite {NC99}.

%example of $q<1$
Consider the example of the discrete part of the spectrum of the
hydrogen atom. It has $q_c=5/3$
(indeed the spectrum consists of eigenvalues $-\alpha/n^2$,
$n=1,2,\cdots$; the $n$-th level is $n^2$-times degenerate;
due to the degeneracy, the
eigenvalues $\epsilon_n$ tend to $\epsilon_\infty=0$ as $-\alpha/n^{2/3}$;
neglect the spin of the electron).
The corresponding dimension equals 3/2 (again, one would
expect the dimension to be equal to 3; the extra factor 1/2 disappears
if the square root of minus the hamiltonian is the operator determining the
geometry). See the figure for a plot of $\alpha\beta_-$ as a
function of $q$ in the domain of existence $(3/5,1)$.

\begin{figure}
\centerline{\psfig{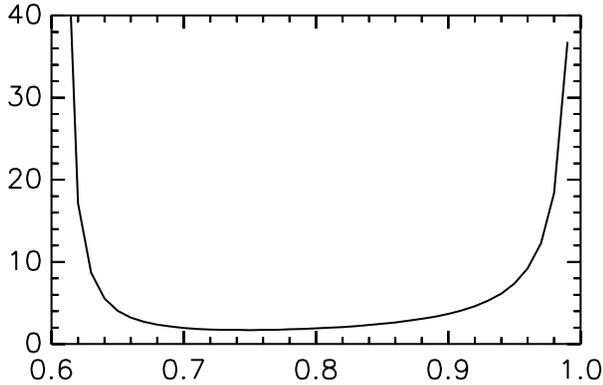}}
\centerline{\parbox[b]{6cm}}
{\caption
{$\alpha\beta_-$ as a function of $q$.}
}
\end{figure}

In \cite {LST95} an anomaly is reported at $q_c=9/7$
instead of $q_c=5/3$. This value was taken from \cite {PP93},
which is a paper
on stellar polytropes. Study of the latter paper makes clear
that it concerns the same anomaly as the one studied here, namely the
upper limit for which $q$-statistics is meaningful.
However, it is not evident that $q_c$ should have the same value in
a classical model of gravitation as in a quantum mechanical model
of Coulomb attraction. Further differences between \cite {LST95}
and the present paper can be explained by the use of a
different expression for the energy functional (\ref {energ}).

One concludes that a statistical description of the hydrogen
atom is possible in the interval $q_c^{-1}<q<1$. The
temperature varies between zero for the ground state and
the maximal value $T_-\equiv1/k_B\beta_-$ which depends itself on $q$.
If the heath bath is very large then $q$ is close to one and the
abundance of energy destabilizes the system (i.e.~$T_-$ tends to
zero). If $q$ tends to $q_c$ then the Tsallis entropy diverges
and again $T_-$ tends to zero.

%Acknowledgements

%

\end{document}